# Temperature and magnetic field dependence of a Kondo system in the weak coupling regime

Yong-hui Zhang[1,2,*], Steffen Kahle[1,*], Tobias Herden[1], Christophe Stroh[3], Marcel Mayor[3,4], Uta Schlickum[1], Markus Ternes[1], Peter Wahl[1,5] & Klaus Kern[1,6]

The Kondo effect arises due to the interaction between a localized spin and the electrons of a surrounding host. Studies of individual magnetic impurities by scanning tunneling spectroscopy have renewed interest in Kondo physics; however, a quantitative comparison with theoretical predictions remained challenging. Here we show that the zero-bias anomaly detected on an organic radical weakly coupled to a Au (111) surface can be described with astonishing agreement by perturbation theory as originally developed by Kondo 60 years ago. Our results demonstrate that Kondo physics can only be fully conceived by studying both temperature and magnetic field dependence of the resonance. The identification of a spin 1/2 Kondo system is of relevance not only as a benchmark for predictions for Kondo physics but also for correlated electron materials in general.

[1] Max-Planck Institute for Solid State Research, Heisenbergstraße 1, 70569 Stuttgart, Germany. [2] Department of Physics, Tsinghua University, Beijing 100084, China. [3] Institute of Nanotechnology, Karlsruhe Institute of Technology, PO Box 3640, 76021 Karlsruhe, Germany. [4] Department of Chemistry, University of Basel, 4056 Basel, Switzerland. [5] SUPA, School of Physics and Astronomy, University of St Andrews, St Andrews KY16 9SS, UK. [6] Institut de Physique de la Matière Condensée, Ecole Polytechnique Fédérale de Lausanne (EPFL), 1015 Lausanne, Switzerland. * These authors contributed equally to this work. Correspondence and requests for materials should be addressed to M.T. (email: m.ternes@fkf.mpg.de) or to P.W. (email: wahl@fkf.mpg.de).

The Kondo effect, discovered originally in dilute magnetic alloys[1], emerges ubiquitously in seemingly unrelated contexts, such as the zero-bias anomalies observed in quantum dots and nanowires[2,3] or the dynamical behavior close to a Mott transition[4,5]. The simplicity of the underlying model Hamiltonian—a single spin coupled by an exchange interaction to a bath of conduction electrons—contrasts the complex physics that only the development of a completely new theoretical understanding clarified[6–8].

The Kondo effect is usually considered for an antiferromagnetic (AFM) interaction between a localized spin and an itinerant spin bath with a spin–spin exchange coupling $J<0$. This interaction leads in the strong coupling regime, that is, at temperatures below a characteristic Kondo temperature, $T_K$, to a screening of the impurity magnetic moment and results in a stable, non-magnetic singlet ground state[9] (see Fig. 1).

Much less attention has been paid to the weak coupling regime, which is relevant for ferromagnetic (FM) interaction ($J>0$) or at elevated temperatures ($T \gg T_K$). For FM interaction, the impurity spin is always weakly coupled and becomes asymptotically free in the limit of low temperature[6,10]. For AFM interactions and high temperatures, thermal fluctuations destroy the singlet state. For both cases, FM interaction and AFM interaction in the weak coupling regime, the physics can be described by perturbation theory[6]. In past studies of single Kondo adsorbates by scanning tunneling spectroscopy, a detailed quantitative characterization of the Kondo physics was hampered by orbital degeneracies[11], spin quantum numbers $>1/2$ (ref. 12), and rather high Kondo temperatures[13,14].

Here, we study a purely organic molecule, which has a radical nitronyl-nitroxide side group[15], adsorbed on a Au (111) surface. Molecules with the same radical side group have been shown to form ferromagnetic molecule crystals below a Curie temperature of 0.6 K (ref. 16). Instead of being localized on a specific atom, the unpaired electron is spatially delocalized over the O–N–C–N–O part of the side group (Fig. 2a) stabilizing it against chemical reaction and charge transfer, which would lead to a spin zero system. The unpaired electron has no further orbital degeneracy.

This enables us to study the physics of a pure spin 1/2 interacting with conduction electrons, which has been the subject of intense theoretical research.

## Results

**Topographic imaging of organic radicals.** Constant-current scanning tunneling microscopy (STM) images acquired at low temperatures show that the molecules adsorbed on the surface tend to nucleate at the elbow sites of the reconstructed Au (111) surface (Fig. 2b). Detailed topographies show the elongated molecular backbone and the radical side group, which is imaged as a ~0.3 nm high and 1.0 nm wide protrusion (Fig. 2c).

**Temperature dependence of the zero-bias peak.** We detect a strong resonance at the Fermi level when measuring the differential conductance (d$I$/d$V$, for details see Methods section) on the side group of the molecule at temperatures between $T=1.5$ K and 15.7 K (Fig. 3a,b), which is found neither on the backbone of the molecule nor on the clean surface. The observation of this resonance does not depend on the orientation of the molecule with respect to the Au (111) substrate or on the presence of codeposited dichloromethane molecules close to the molecule, while its apparent width and amplitude varies slightly between different molecules. The resonance shows clear logarithmic voltage dependence over almost two orders of magnitude (Fig. 3b) and broadens significantly at increased temperatures—much more than expected for a single-particle resonance. The apparent width of the resonance does not converge to a natural width in the low-temperature limit within the temperature range accessible to us, as one would expect for a spin 1/2 Kondo system with AFM interaction in the strong coupling limit at $T \ll T_K$. The spectra rather have a flat top whose width scales with the thermal energy $k_B T$ ($k_B$ is the Boltzmann constant).

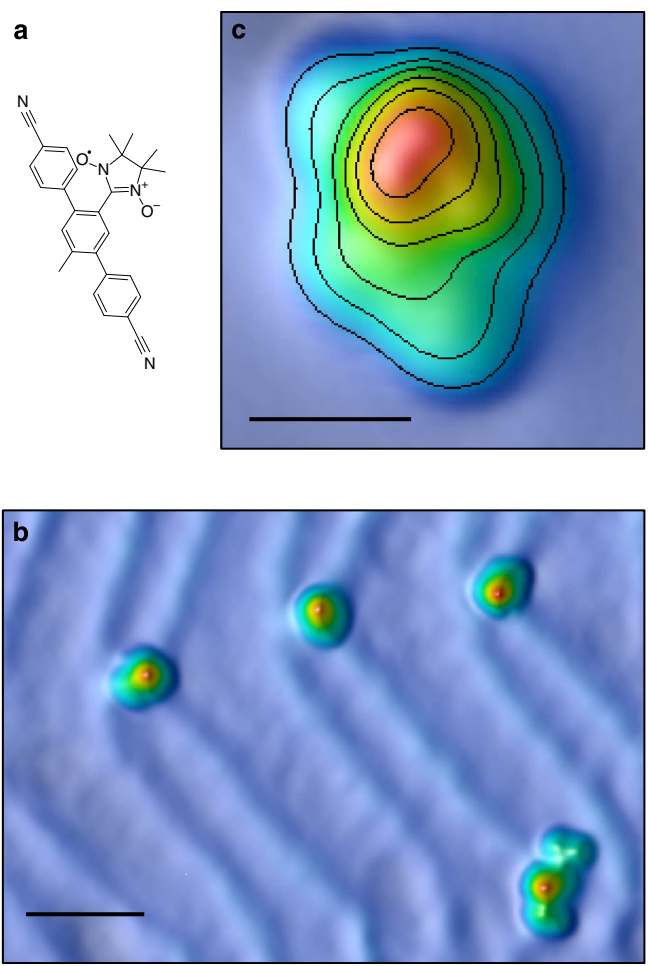

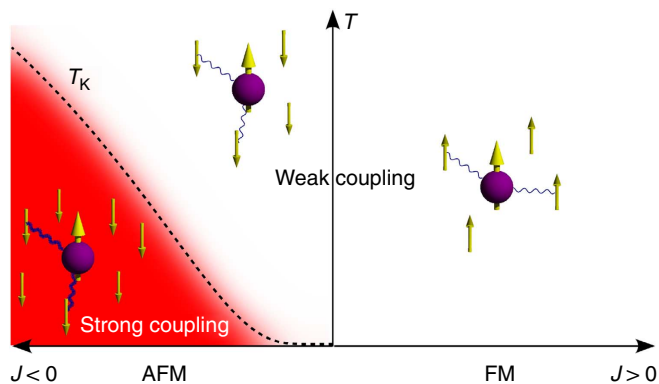

**Figure 1 | The different regimes of the Kondo effect.** The Kondo interaction $-J\,\mathbf{S}\cdot\boldsymbol{\sigma}$ couples itinerant electrons of the host with spin $\boldsymbol{\sigma}$ to a magnetic impurity with spin $\mathbf{S}$. For exchange interaction $J<0$, the antiferromagnetic (AFM) coupling leads to an entangled many-body state, where the antiparallel alignment of the spins of the conduction electrons effectively screens the impurity spin. The ground state at temperatures $T$ below the characteristic Kondo temperature $T_K$ is a singlet with total spin $S=0$ (red area), well protected from higher energy states. In contrast, for $J>0$, the ferromagnetic (FM) coupling tends to create a screening cloud of spins aligned parallel to the impurity spin, which becomes asymptotically free at low temperatures. For FM coupling or at temperatures $T \gg T_K$, the system is in the weak coupling regime, which can be treated perturbatively.

**Figure 2 | Topography of an organic radical molecule on Au (111).** (**a**) Chemical structure of the studied organic radical molecule ($C_{28}H_{25}O_2N_4$) with a nitronyl-nitroxide side group that contains a delocalized singly occupied molecular orbital. The molecule is drawn with a similar orientation as in the topography in **c**. (**b**) Overview topography ($T=6.7$ K, bias $V=50$ mV, setpoint current $I=30$ pA) of individual molecules on Au(111). The bottom right molecule is decorated by three neighboring dichloromethane molecules. The scale bar corresponds to 5 nm. (**c**) High-resolution topography of one organic radical molecule ($V=100$ mV, $I=33$ pA). Contour lines are at height intervals of 50 pm. The scale bar corresponds to 1 nm.

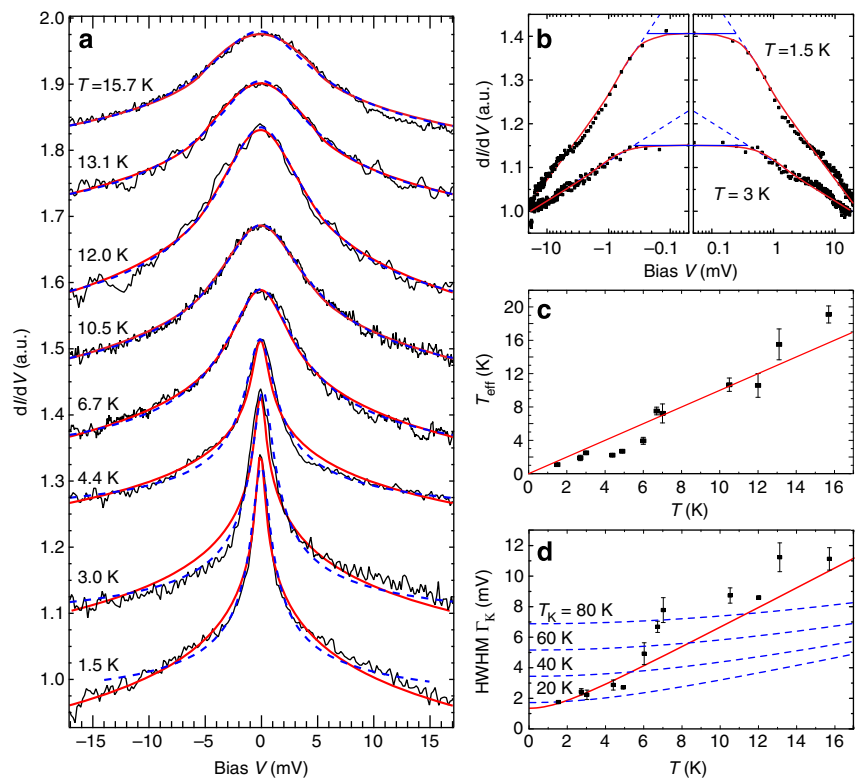

**Figure 3 | Temperature dependence of the zero-bias anomaly.** (**a**) Typical differential conductance spectra taken on the radical side group of the molecule at temperatures between $T=1.5$ K and $T=15.7$ K (black), simulated spectra using perturbation theory (red) and fits of a Frota function (blue dashed lines). All spectra are normalized and offset for visual clarity. (**b**) Two exemplary spectra (black dots) and the conductance obtained from the Anderson–Appelbaum model (red line) plotted with logarithmic abscissa. The blue lines mark the flat top range (intersection between conductance at zero bias and an extrapolation of the logarithmic increase in the conductance, shown by dashed lines) whose width scales with temperature. (**c**) Plot of the effective temperature $T_{\text{eff}}$ obtained by fitting spectra of about 35 molecules using the Anderson–Appelbaum model versus the experimental temperature $T$. The solid line is $T_{\text{eff}}=T$. Error bars include the variation from spectra taken on different molecules. (**d**) Half-width at half-maximum (HWHM) extracted from the fits of Frota functions in **a** as function of temperature, the blue dashed lines indicate the expected temperature dependence of the width for a Kondo resonance in the strong coupling regime in Fermi liquid theory $\Gamma_K = \frac{1}{2}\sqrt{(\alpha k_B T)^2 + (2k_B T_K)^2}$ with $\alpha = 2\pi$ (ref. 31), the red line is a fit with free $\alpha = 15.2$.

We compare our data with the conductance calculated from the Kondo spin-flip scattering Hamiltonian[1] in a perturbative approach accounting for processes up to third order in the exchange interaction $J$. In this Anderson–Appelbaum model[17–19], tunneling between two electrodes via a magnetic impurity leads at zero magnetic field to a temperature-broadened logarithmic resonance in the conductance as a function of bias, $\epsilon = eV$ ($e$ is the elementary charge), of the form

$$\sigma(\epsilon) = -(J\rho_0)^3 \int_{-\omega_0}^{\omega_0} \frac{f(\epsilon',T)}{\epsilon - \epsilon'} d\epsilon' * f'(\epsilon,T) + c \quad (1)$$

where the asterisk denotes a convolution, $f(\epsilon,T)$ is the Fermi–Dirac distribution, $f'(\epsilon,T)$ its derivative with regard to $\epsilon$, $\rho_0$ the density of states of the conduction band at the Fermi energy, $\omega_0$ a cutoff energy and $c$ a constant background. While the perturbative approach might have some limitations, more recent approaches that treat non-equilibrium weak coupling Kondo effects more accurately[20] would need generalization to finite temperatures to be applicable here. It has been shown that the perturbative approach gives a correct description if the magnetic impurity is coupled predominantly to one of the two electrodes and the system is close to equilibrium[21].

We perform least-squares fits, taking the temperature as the only relevant parameter (for details see Supplementary Note 1). We denote the temperature obtained from the fits $T_{\text{eff}}$. Figure 3a,b show the excellent agreement between the model and our data. The extracted effective temperatures, $T_{\text{eff}}$, agree well with the temperature $T$ of the experiment with deviations towards lower values for $T_{\text{eff}}$ at temperatures $T \leq 6$ K (Fig. 3c). For comparison, in Fig. 3a, also fits of a Frota function[22] to the data are shown, with the extracted half-widths plotted in Fig. 3d.

**Magnetic field dependence of spectra.** When we apply a magnetic field $B$ perpendicular to the Au (111) surface, the resonance splits into two peaks with superimposed symmetric steps as soon as the Zeeman energy exceeds the thermal energy (Fig. 4a), that is, $g_0 \mu_B B > k_B T$ ($\mu_B$ is the Bohr magneton, $g_0$ the Landé factor of a free electron). The steps in the differential conductance are due to inelastic spin-flip excitations[23] and can be described with excellent agreement using the same perturbative approach (for details see Supplementary Note 1)[24]. The perturbation theory not only accurately accounts for the steps, but also for the peaks on top of these steps. The energies of the peak positions and the spin-flip excitations as extracted from the fits scale linearly with the applied magnetic field and yield a Landé factor $g = 1.98 \pm 0.04$, which slightly differs from the expected value for a free electron, $g_0 = 2$. This difference provides an estimate for the exchange interaction $J\rho_0$: due to the coupling of the localized spin with the conduction electrons and their polarization by the magnetic field, the effective Landé factor will be modified compared to the value for the free spin[25]. For an evaluation of $J\rho_0$, the error in the determination of the peak and step positions and hence of $g$ is critical (see Supplementary Figure S2). As can be seen from Fig. 4b,c, the error bars become smaller for larger magnetic field. This is because only at large magnetic fields, the spectra become flat around zero-bias voltage, increasing the accuracy of the fits. Therefore, to estimate $J\rho_0$ from the Zeeman splitting, we evaluate $g$ only for the highest magnetic fields ($B \geq 10$ T). We obtain $g = 1.93 \pm 0.02$ and $J\rho_0 = -0.04 \pm 0.01$. We derive a second independent estimate of $J\rho_0$ by comparing the amplitudes of the different scattering orders in the model: The height of the logarithmic peaks is proportional to $(J\rho_0)^3$, while the amplitude of the steps due to spin-flip excitations is proportional to $(J\rho_0)^2$. We find $J\rho_0 = -0.04 \pm 0.02$, in agreement with the value obtained from the Zeeman splitting (Fig. 4c). Our fits indicate a weak AFM exchange interaction between the localized spin on the molecule and the Au surface. FM coupling between the spin and the conduction electrons as depicted in Fig. 1 would lead to similar physics. The spectroscopic signature for an FM exchange interaction is a dip in the spectral function at the impurity site, while in the conduction band of the host, a peak at the Fermi energy is formed[26]—opposite to the case of AFM coupling. This result can also be obtained from perturbation theory, which for FM coupling remains valid at all temperatures, as the sign of $g(\epsilon)$ changes for $J > 0$ (see equation 1). In our measurements, we always detect a peak at the radical side group. From this finding and the estimation of $J\rho_0$, we can exclude FM coupling in our system. In contrast to the case of adatoms, where the Kondo resonance is assumed to be probed rather indirectly through the conduction band of the surface[27], here the resonance is probed directly by tunneling from the tip into the many body states in accordance with the Anderson–Appelbaum theory. For a discussion of the case of weak coupling of the tip to the adsorbate resonance see Supplementary Note 3.

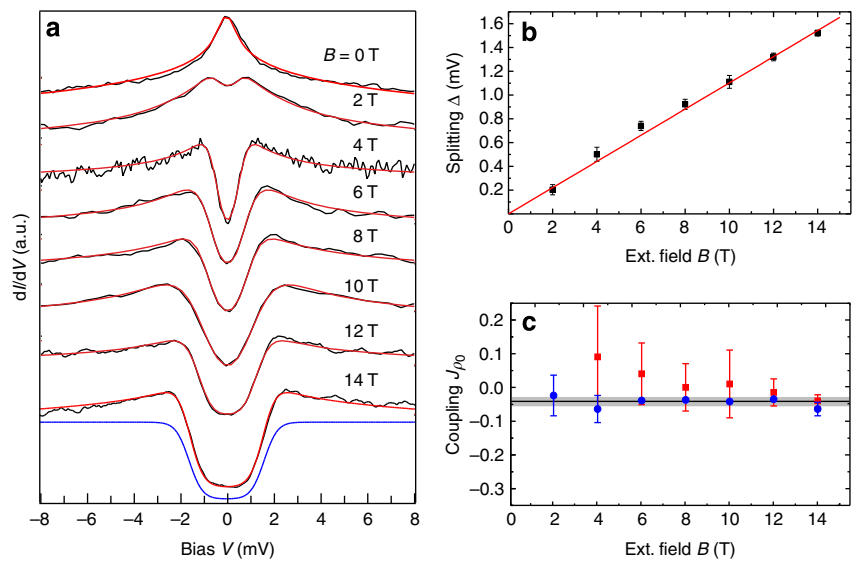

**Figure 4 | Splitting of the resonance in magnetic field.** (**a**) Differential conductance measurements taken at successively increased magnetic fields on the radical side group of the molecule ($T = 1.8$ K). All spectra are normalized and offset for clarity. Red curves are fits modeling the conductance using perturbation theory up to third order in the exchange interaction $J$. The blue curve at $B = 14$ T shows exemplarily the contribution of the second order in $J$ to the differential conductance. (**b**) extracted Zeeman splitting as a function of magnetic field. A linear fit (red line) yields $g = 1.98 \pm 0.04$. (**c**) magnetic coupling $J\rho_0$ obtained from the ratio between second and third order contributions to the fit (blue dots), with an average coupling of $J\rho_0 = -0.04 \pm 0.02$ and from the Zeeman splitting (and hence the $g$ factor) (red dots). For the latter analysis, at each magnetic field the Zeeman splitting $\Delta_Z(B)$ is determined from the fits (see also Supplementary Note 1) and the $g$ factor is obtained from $g(B) = \Delta_Z(B)/(\mu_B B)$. Error bars in **b** and **c** include the variation of spectra taken on different molecules.

## Discussion

In the past, the zero-bias anomalies observed by scanning tunneling spectroscopy on single atoms and molecules have usually been interpreted as a Kondo resonance in the strong coupling limit. Analyzing our data with this assumption by fitting the spectra by a Frota function[22], which has been shown to describe well the Kondo resonance (at $T < T_K$)[22,28], yields at first glance a reasonable description of our data (see Fig. 3a; Supplementary Note 2). In the low-temperature limit, we obtain a half-width of $\Gamma_K \approx 1.3$ mV, which would correspond to $T_K = \Gamma_K/k_B \approx 16$ K. The magnetic field dependence differs from the behavior expected for an AFM Kondo model for a spin 1/2 in the strong coupling limit[29,30] and a $T_K \approx 16$ K. For $T \ll T_K$, the Kondo resonance is not expected to split for magnetic fields up to $B_C \approx 0.5 k_B T_K/(g\mu_B)$ (ref. 29). Thus, we should observe a splitting only at $B > B_C \approx 6$ T, contrary to our observation of a clear splitting at fields as low as 2 T. This yields for our system an upper limit of $T_K \lesssim 5.4$ K. Furthermore, there should be only a rather weak temperature dependence of the width of the resonance for $T < T_K$ (ref. 31), which would be the temperature range of our measurements, contrary to our observations (see Fig. 3d).

The disagreement arises because the half-width of a logarithmic singularity is not a well-defined quantity and thus the determination of $T_K$ via the half-width is erroneous. Only by taking into account both, the temperature and magnetic field dependence, we can deduce that the true Kondo temperature must be substantially lower.

We conclude that the resonance we observe is due to a single unpaired spin with weak AFM coupling to the conduction band. Almost 60 years after Kondo identified the correct Hamiltonian for a single magnetic impurity in a metallic host and showed the breakdown of perturbation theory on a characteristic energy scale[1], our results demonstrate its validity for a single spin 1/2 impurity in the weak coupling regime, which is characterized by an almost universal temperature and magnetic field dependence. Both differ substantially from the behavior predicted for a spin 1/2 AFM Kondo model in the strong coupling limit. Even though zero-bias anomalies have been observed previously on purely organic molecules adsorbed on metal surfaces[32,33], the quantitative level of agreement between the data and theory is unprecedented for measurements on single impurities by low-temperature STM.

The identification of a true spin 1/2 system allows to perform quantitative tests of theoretical predictions, which can be tested in measurements performed at even lower temperatures. Besides enabling studies of spin 1/2 Kondo physics on a single impurity, experiments on systems of coupled impurities are conceivable. Here, we note that most theoretical studies of Kondo lattices are built on top of spin 1/2 Kondo impurities. Tailoring of the backbone in the synthesis of the molecules facilitates self-assembled growth of ordered one- or two-dimensional lattices of these molecules. This potentially allows for experimental studies of model Hamiltonians for correlated electron materials.

## Methods

**STM measurements.** Experiments were performed on two different home-built STMs operating in ultra-high vacuum, one with a base temperature of 6.7 K and magnetic fields up to 5 T, the second one with a base temperature of 1.5 K and fields up to 14 T. Both setups allow *in situ* sample preparation and transfer into the STM.

**Differential conductance spectroscopy.** Conductance spectra were acquired by positioning the tip over the side group of the molecule, turning the z-feedback off, and sweeping the bias voltage $V$. To detect the d$I$/d$V$ signal directly by lock-in technique, we add to $V$ a small sinusoidal modulation voltage $V_m$ with a frequency in the range of 800–1,100 Hz. Typical tunneling setpoints before deactivating the feedback loop were $V = -20$ mV and $I = 0.1$–1 nA. Spectra were typically taken in a range of $-20$ mV to $+20$ mV symmetrically around zero bias. $V_m$ was chosen to be small enough that the signal broadening due to the modulation voltage was always less than the thermal broadening; more precisely $V_m(\text{rms}) < \frac{1}{2} 5.4 k_B T$, resulting in typical used values of $V_m(\text{rms}) = 0.1$–0.3 mV at low temperatures, while at higher temperatures ($T > 5$ K) 0.5–1 mV was used. Before and after spectra were measured on the radical, reference spectra have been taken on the clean Au surface to ensure a featureless conductance of the tip. The measured spectra were normalized and a linear baseline was subtracted to account for drift effects that were also visible in the reference spectra.

**Sample preparation.** The Au (111) surface is prepared by cycles of Ar$^+$ sputtering and annealing to $\sim 800$ K. Molecules are deposited from crystal powder containing equal amounts of the radical molecule and dichloromethane by sublimation using a quartz crucible at a temperature of $\approx 200$ °C for several minutes with the sample at $T = 150$ to 180 K. It has been confirmed by matrix-assisted laser desorption/ionization that the molecules stay intact during sublimation. The molecules are only weakly bound to the surface; deposition at room temperature leads to nucleation at step edges.

**Synthesis.** The molecule (2'-nitronilnitroxide-5'-methyl-[1, 1';4',1'']terphenyl-4,4''-dicarbonitrile ($C_{28}H_{25}O_2N_4$)) has been synthesized using a Suzuki-type cross-coupling reaction with the spin-labeled diiodo-precursor and 4-cyano-phenyl-boronic acid in the presence of Pd(PPh$_3$)$_4$ as catalyst and Na$_2$CO$_3$ as base.

## Acknowledgements

We gratefully acknowledge intense discussions and sharing of unpublished results by P.P. Baruselli, R. Requist, M. Fabrizio, and E. Tosatti. Further, we acknowledge discussions with D. Le, and T. Rahman, and with T. Costi, C. Hooley, N. Lorente, J. Merino, E. Sela, P. Simon and R. Zitko. We thank S. Rauschenbach for performing MALDI analysis on test depositions. Y.Z. acknowledges support by the Chinese Scholarship Council. S.K., T.H., M.T. and P.W. acknowledge support by SFB 767 and US by the Emmy-Noether-programme. Support by the Baden-Württemberg-Stiftung is acknowledged. CS and MM acknowledge financial support by the Alexander-von-Humboldt Foundation.

### Author contributions

P.W., M.T. and K.K. conceived and supervised the project; Y.Z., S.K. and T.H. performed experiments with help from U.S.; C.S. and M.M. synthesized and purified the molecules; M.T., P.W., S.K. and Y.Z. did data analysis; all authors contributed to and discussed the manuscript.

### Additional information

**Supplementary Information** accompanies this paper at http://www.nature.com/naturecommunications

**Competing financial interests:** The authors declare no competing financial interests.

**Reprints and permission** information is available online at http://npg.nature.com/reprintsandpermissions/

**How to cite this article:** Zhang, Y.-h. *et al.* Temperature and magnetic field dependence of a Kondo system in the weak coupling regime. *Nat. Commun.* 4:2110 doi: 10.1038/ncomms3110 (2013).

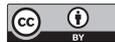